\documentclass[12pt]{article}
\begin{document}

\begin{center}
{\Large\bf A modification of the Chen-Nester quasilocal expressions}
\end{center}

\begin{center}
Lau Loi So \\
Department of Physics, National Central University, Chung-Li 320,
Taiwan.
\end{center}

\begin{abstract}
Chen and Nester proposed four boundary expressions for the
quasilocal quantities using the covariant Hamiltonian formalism.
Based on these four expressions, there is a simple generalization
that one can consider, so that a two parameter set of boundary
expressions can be constructed.  Using these modified expressions, a
nice result for gravitational energy-momentum can be obtained in
holonomic frames.
\end{abstract}

\section{Introduction}
For any field theory, Chen and Nester~\cite{Nester,Nester 1}
proposed certain boundary expressions for the quasilocal quantities.
They found four special expressions.  Two are unique under the
property of their three requirements: the well defined requirement,
the symplectic structure requirement and the covariant requirement.
The other two correspond to boundary conditions imposed on a
physically meaningful but non-covariant set of variables.
Generalizing the latter, in the present paper we make some
modification by a simple adjustment of the four boundary
expressions.  Using these four expressions as the basis, one can
obtain a two parameter set of boundary expressions.  The modified
two parameter set of expressions for general relativity includes one
case with a nice \symbol{92}positive energy'' result in holonomic
frames in small regions.

\section{Chen-Nester's 4 boundary expressions}
Following~\cite{Nester}, begin with a first order Lagrangian density
(i.e., a 4-form)
\begin{equation}\label{18Mar2006}
{\cal{}L}=dq\wedge{}p-\Lambda(q, p),
\end{equation}
where $q$ and $p$ are the canonical conjugate form fields, $\Lambda$
is a potential, we suppose that $q$ is a $f$-form and let
$\epsilon=(-1)^{f}$.  The corresponding Hamiltonian 3-form is
defined as
\begin{equation}
{\cal{}H}(N):=\pounds_{N}q\wedge{}p-i_{N}{\cal{}L}.
\end{equation}
Taking the interior product of the Lagrangian density
(\ref{18Mar2006})
\begin{eqnarray}
i_{N}{\cal{}L}&=&i_{N}dq\wedge{}p-\epsilon{}dq\wedge{}i_{N}{}p-i_{N}\Lambda
\nonumber\\
&=&\pounds_{N}q\wedge{}p
-\epsilon{}i_{N}q\wedge{}dp-\epsilon{}dq\wedge{}i_{N}p
-i_{N}\Lambda-d(i_{N}q\wedge{}p),
\end{eqnarray}
where the Lie derivative or form component is
$\pounds_{N}:=i_{N}d+di_{N}$. We see that the Hamiltonian 3-form
(density) can be put in the form
\begin{equation}
{\cal{}H}(N)=N^{\mu}{\cal{}H}_{\mu}+d{\cal{}B}(N),
\end{equation}
where
\begin{equation}\label{19}
N^{\mu}{\cal{}H}_{\mu}=i_{N}q\wedge{}dp
+\epsilon{}dq\wedge{}i_{N}p+i_{N}\Lambda,
\end{equation}
and the natural boundary term is
\begin{equation}
{\cal{}B}(N)=i_{N}q\wedge{}p.
\end{equation}
This is called the boundary expression, because when one integrates
the Hamiltonian density over a finite region to get the Hamiltonian,
the boundary term leads to an integral over the boundary of the
region. However this boundary term is not unique, one can add to it
whatever he likes without changing the Hamilton equations. Indeed
this boundary term can be removed by introducing the new Hamiltonian
3-form as follows
\begin{equation}
{\cal{}H'}(N)={\cal{}H}(N)+d(-i_{N}q\wedge{}p)
=N^{\mu}{\cal{}H}_{\mu}.
\end{equation}
The variation of the Hamiltonian (\ref{19}) has the form (for
details see~\cite{Nester 2})
\begin{equation}
\delta{\cal{}H'}(N)=-i_{N}({\rm{}F.E.}) -\delta{}q\wedge\pounds_{N}p
+\pounds_{N}q\wedge\delta{}p+d{\cal{}C}(N),
\end{equation}
where the field equation term
\begin{equation}
{\rm{}F.E.}:=\delta{}q\wedge\frac{\delta{\cal{}L}}{\delta{}q}
+\frac{\delta{\cal{}L}}{\delta{}p}\wedge\delta{}p,
\end{equation}
is proportional to the first order Lagrangian field equations and
can be assumed to vanish.   The boundary variation term is
\begin{equation}\label{16}
{\cal{}C}(N)=-i_{N}q\wedge\delta{}p
+\epsilon\delta{}q\wedge{}i_{N}p.
\end{equation}
This boundary variation term cannot be removed because it comes from
$\delta{\cal{}H'}$ directly.  Boundary conditions are obtained by
requiring that the boundary term in the variation of the Hamiltonian
vanish.  We have to add a suitable boundary term to the Hamiltonian
3-form (\ref{19})
\begin{equation}
{\cal{}H'}(N)\rightarrow{\cal{}H}_{k}(N)
=N^{\mu}{\cal{}H}_{\mu}+d{\cal{}B}_{k}(N),
\end{equation}
to modify the variational boundary term, in order to get nice
components like $i_{N}(\delta{}q\wedge\Delta{}p)$ or
$i_{N}(\Delta{}q\wedge\delta{}p)$. In order to achieve such forms,
there are just four simple boundary expressions that can be added,
as can be seen by referring to (\ref{16}). The variation of the four
Hamiltonians including these different boundary expressions are
\begin{eqnarray}
\delta{\cal{}H}_{q}(N)&=&K +di_{N}(\delta{}q\wedge\Delta{}p)
=K+d(i_{N}\delta{}q\wedge\Delta{}p
+\epsilon\delta{}q\wedge{}i_{N}\Delta{}p),\label{17} \\
\delta{\cal H}_{p}(N)&=&K-di_{N}(\Delta{}q\wedge\delta{}p)
=K-d(i_{N}\Delta{}q\wedge\delta{}p
+\epsilon\Delta{}q\wedge{}i_{N}\delta{}p), \\
\delta{\cal{}H}_{\rm{}d}(N)&=&K +d(-i_{N}\Delta{}q\wedge\delta{}p
+\epsilon\delta{}q\wedge i_{N}\Delta p), \\
\delta{\cal{}H}_{\rm{}c}(N)&=&K +d(i_{N}\delta{}q\wedge\Delta p
-\epsilon\Delta{}q\wedge i_{N}\delta p),\label{18}
\end{eqnarray}
where $K$ is defined as
\begin{equation}
K:=-i_{N}({\rm{}F.E.})
-\delta{}q\wedge\pounds_{N}p+\pounds_{N}q\wedge\delta{}p.
\end{equation}
Thus we recover the Chen-Nester 4 boundary expressions
\begin{eqnarray}
{\cal B}_{q}(N)&=&i_{N}q\wedge\Delta
p-\epsilon\Delta{}q\wedge i_{N}\overline{p}, \\
{\cal B}_{p}(N)&=&i_{N}\overline{q}\wedge\Delta
p-\epsilon\Delta{}q\wedge{}i_{N}{p}, \\
{\cal B}_{\rm d}(N)&=&i_{N}\overline{q}\wedge\Delta p
-\epsilon\Delta{}q\wedge i_{N}\overline{p}, \\
{\cal B}_{\rm c}(N)&=&i_{N}q\wedge\Delta p -\epsilon\Delta{}q\wedge
i_{N}{p},
\end{eqnarray}
(refer to \cite{Nester 1,Nester 2} for an explanation of the
terminology) where ${\cal B}_{\rm{}d}$ stands for ${\cal
B}_{\rm{}dynamics}$, ${\cal B}_{\rm{}c}$ for ${\cal
B}_{\rm{}constraint}$, here $\Delta{}q=q-\overline{q}$,
$\Delta{}p=p-\overline{p}$, $\overline{q}$ and $\overline{p}$ are
the background reference values.  Note that the above four boundary
expressions can be written in a compact form
\begin{equation}
{\cal B}_{k_{1},k_{2}}(N)={\cal B}_{p}(N)
+k_{1}i_{N}\Delta{}q\wedge\Delta{}p+\epsilon
k_{2}\Delta{}q\wedge{}i_{N}\Delta{}p,
\end{equation}
where $(k_{1},k_{2})=(0,0),(0,1),(1,0)$ and $(1,1)$.  In detail
\begin{eqnarray}
{\cal{}B}_{0,0}={\cal B}_{p}, \quad
{\cal{}B}_{0,1}={\cal{}B}_{\rm{}d},\quad
{\cal{}B}_{1,0}={\cal{}B}_{\rm{}c}, \quad
{\cal{}B}_{1,1}={\cal{}B}_{q}.
\end{eqnarray}
The boundary conditions associated with ${\cal{}B}_{q}$ and
${\cal{}B}_{p}$ are the simplest. For the case of ${\cal{}B}_{q}$,
we want the variation boundary term to satisfy
$i_{N}(\delta{}q\wedge\Delta{}p)=0$. There are two ways to achieve
this requirement.  First control $q$, then $\delta{}q=0$. The second
is to freely vary $q$ and then $\delta{}q$ can be anything which
implies that we want $\Delta{}p=0$. This is the {\symbol{92}}natural
boundary condition", it forces $p=\overline{p}$, where
$\overline{p}$ is the background reference.  Similarly for the
variation boundary term
$i_{N}(\Delta{}q\wedge\delta{}p)=0$.\\

A simple example of the idea is the one dimensional spring with a
mass block, control the external force $\delta{}F$ and the length of
the spring $\Delta{}x$ changes or response. Likewise, control the
position of the mass block by $\delta{}x$, there must be a force
$\Delta{}F$ from outside the system as response. However the
variation boundary expressions from ${\cal{}B}_{{\rm{}d}}$ and
${\cal{}B}_{\rm{}c}$ are more complicated, because they both have
mixed control-$q$, $p$-response and control-$p$, $q$-response type
terms at the same time.

\section{The modification of the Chen-Nester 4 boundary expressions}
Consider equation (\ref{17}), as long as the variational boundary
terms pieces $i_{N}\delta{}q\wedge\Delta{}p$ and
$\delta{}q\wedge{}i_{N}\Delta{}p$ vanish.  It seems that there is no
restriction on the magnitude.  In order words, it can be any real
number that one can impose.  Consider the variation of the boundary
expression in a simple way
\begin{equation}
\delta\widetilde{{\cal{}H}}_{q}
=K+d(k_{1}i_{N}\delta{}q\wedge\Delta{}p
+\epsilon{}k_{2}\delta{}q\wedge{}i_{N}\Delta{}p).
\end{equation}
However, one could construct other similar forms proceeding from
$\delta{\cal{}H}_{p}$, $\delta{\cal{}H}_{\rm{}d}$ or
$\delta{\cal{}H}_{\rm{}c}$.  The general form of the variational
Hamiltonian of the four different expressions, the modified
equations (\ref{17}) to (\ref{18}) in one general form all together
is
\begin{equation}
\delta\widetilde{{\cal{}H}}
=K+b_{1}i_{N}\delta{}q\wedge\Delta{}p+b_{2}\delta{}q\wedge{}i_{N}\Delta{}p
+b_{3}i_{N}\Delta{}p\wedge\delta{}q+b_{4}\Delta{}p\wedge{}i_{N}\delta{}q,
\end{equation}
where $b_{1}$ to $b_{4}$ are real numbers (it will turn out that
they cannot be all independent). All of these extra terms can be
obtained by adding a suitable multiple of
$i_{N}\Delta{}q\wedge\Delta{}p$ or
$\Delta{}q\wedge{}i_{N}\Delta{}p$. Actually one can rewrite the
modified Chen-Nester 4 boundary expressions in a simple way by
adding
$c_{1}i_{N}\Delta{}q\wedge\Delta{}p+\epsilon{}c_{2}\Delta{}q\wedge
i_{N}\Delta{}p$ for the individual boundary expressions, where
$c_{1},c_{2}\in\Re$. In detail, the 4 equivalent forms are
\begin{eqnarray}
\widetilde{\cal{}B}_{q}&=&i_{N}q\wedge\Delta{}p-\epsilon\Delta{}q\wedge
i_{N}\overline{p}+c_{1}i_{N}\Delta{}q\wedge\Delta
p+\epsilon{}c_{2}\Delta{}q\wedge{}i_{N}\Delta{}p, \label{1} \\
\widetilde{\cal{}B}_{p}&=&i_{N}\overline{q}\wedge\Delta{}p-\epsilon\Delta{}q\wedge
i_{N}{p}+c_{1}i_{N}\Delta{}q\wedge\Delta p+\epsilon
c_{2}\Delta{}q\wedge{}i_{N}\Delta{}p, \label{1a}\\
\widetilde{\cal B}_{\rm d}&=&i_{N}\overline{q}\wedge\Delta{}p
-\epsilon\Delta{}q\wedge
i_{N}\overline{p}+c_{1}i_{N}\Delta{}q\wedge\Delta
p+\epsilon{}c_{2}\Delta{}q\wedge
i_{N}\Delta{}p, \label{1b}\\
\widetilde{\cal{}B}_{\rm{}c}&=&i_{N}q\wedge\Delta{}p
-\epsilon\Delta{}q\wedge{}i_{N}{p}+c_{1}i_{N}\Delta{}q\wedge\Delta{}p
+\epsilon{}c_{2}\Delta{}q\wedge{}i_{N}\Delta{}p. \label{4}
\end{eqnarray}
They are equivalent because a suitable choice for $c_{1}$, $c_{2}$
in (\ref{1}) can reproduce (\ref{1a}) to (\ref{4}) with new values
for $c_{1}$, $c_{2}$.  For example in (\ref{1}), when
$c_{1}\rightarrow{}(c_{1}-1)$ gives (\ref{1b}).  In short, one can
rewrite the above four expressions in a compact form
\begin{equation}\label{8Feb2006}
{\cal{}B}_{c_{1},c_{2}}(N)={\cal{}B}_{p}(N)
+c_{1}i_{N}\Delta{}q\wedge\Delta p+\epsilon
c_{2}\Delta{}q\wedge{}i_{N}\Delta{}p.
\end{equation}
If the modification can be adjusted by the above 2 parameters of
$c_{1}$ and $c_{2}$, then what is the role they are playing? Let's
consider the asymptotic order magnitude of the variables $q$,
$\overline{q}$, $p$ and $\overline{p}$ as follows
\begin{eqnarray}
q\approx{}O(1)+O\left(\frac{1}{r}\right),\quad
\overline{q}\approx{}O(1), \quad
p\approx{}O\left(\frac{1}{r^2}\right),\quad \overline{p}\approx{}0.
\end{eqnarray}
Since
\begin{equation}
\Delta{}q\wedge\Delta{}p\approx{}O\left(\frac{1}{r^{3}}\right),
\end{equation}
then the terms with the coefficients of $c_{1}$ and $c_{2}$
asymptotically have $O(1/r^{3})$ fall off.  Consequently they do not
change the dominate main asymptotic value.  This shows that the
modified expression of $c_{1}i_{N}\Delta{}q\wedge\Delta
p+\epsilon{}c_{2}\Delta{}q\wedge i_{N}\Delta p$ adjusts the higher
order terms; but not the lowest order terms, the dominate
contribution.  Indeed the modified expressions are really doing a
modified job.  Consider the variation of the Hamiltonians
\begin{eqnarray}
\delta\widetilde{\cal{}H}_{q}=K+d[(c_{1}+1)i_{N}\delta{}q\wedge\Delta{}p
+\epsilon(c_{2}+1)\delta{}q\wedge{}i_{N}\Delta{}p
+c_{1}i_{N}\Delta{}q\wedge\delta{}p+\epsilon{}c_{2}\Delta{}q\wedge
{}i_{N}\delta{}p],\\
\delta\widetilde{\cal{}H}_{p}
=K+d[c_{1}i_{N}\delta{}q\wedge\Delta{}p,
+\epsilon{}c_{2}\delta{}q\wedge{}i_{N}\Delta{}p
+(c_{1}-1)i_{N}\Delta{}q\wedge\delta{}p
+\epsilon(c_{2}-1)\Delta{}q\wedge{}i_{N}\delta{}p], \\
\delta\widetilde{\cal{}H}_{\rm{}d}
=K+d[c_{1}i_{N}\delta{}q\wedge\Delta{}p+\epsilon{}
(c_{2}+1)\delta{}q\wedge{}i_{N}\Delta{}p
+(c_{1}-1)i_{N}\Delta{}q\wedge\delta{}p+\epsilon
c_{2}\Delta{}q\wedge{}i_{N}\delta{}p], \\
\delta\widetilde{\cal{}H}_{\rm{}c}
=K+d[(c_{1}+1)i_{N}\delta{}q\wedge\Delta{}p +\epsilon
c_{2}\delta{}q\wedge{}i_{N}\Delta{}p
+c_{1}i_{N}\Delta{}q\wedge\delta{}p+\epsilon{}
(c_{2}-1)\Delta{}q\wedge{}i_{N}\delta{}p].
\end{eqnarray}
Basically all of these four boundary expressions both have terms of
the form control-$q$, $p$-response and control-$p$, $q$-response.
The boundary conditions associated with these four expressions are
somewhat like the expressions of ${\cal{}B}_{\rm{}d}$ and
${\cal{}B}_{\rm{}c}$.

\section{On the physical interpretation of the modified
Chen-Nester's 4 boundary expressions} The modified expressions look
a little strange, is this just a mathematical completeness game or
they do have some real physical contribution? What is the idea here?
Does it have any application?  In the Sturm-Liouville theory of 2nd
order linear differential equation. The necessary condition for
self-adjoint is (see e.g. \cite{Arfken})
\begin{equation}
p(y_{1}y'_{2}-y_{2}y'_{1})|_{a}^{b}=0,
\end{equation}
if $p(a)\neq{}0\neq{}p(b)$ one needs to impose boundary conditions
on the solutions.  The simplest is Dirichlet $y(a)=0=y(b)$ or
Neumann $y'(a)=0=y'(b)$.  But a more general choice is to take some
linear combination $\alpha{}y+\beta{}y'$ to vanish on the boundary.
This is an example of the sort of thing we are considering.  Such a
boundary condition is appropriate for some practical applications
such as an elastic cord which is held and shook. Then the slope at
the end point will be proportional to the
end point amplitude.\\

But when we consider field theory, the idea seems not very suitable.
For example for an electrostatic system, such as a parallel plate
capacitor, we know how to physically impose Dirichlet or Neumann
boundary conditions, by fixing the voltage or the charge density,
but it seems not at all simple to construct a device to fix some
linear combination of the voltage and charge density, and it is hard
to imagine why one would want to so such a
thing.\\

Turning from the well understood electrodynamics boundary problem to
the not yet so well understood gravitational field boundary value
problem, we might at first think that the new combined boundary
condition expressions would have little use.  However a little more
thought suggests a different view.  Unlike the electrodynamic case
we do not know physically how to arrange for Dirichlet or Neumann
boundary conditions.  When I imagine trying to change the metric on
the boundary by moving a mass outside, it seems that there will be
associated changes in the normal derivative of the metric. Maybe it
is physically easier to fix a combination of the metric and its
derivatives.  In any case, we found an interesting analytic
application for our idea to gravitational energy.

\section{Gravitation application of the new quasilocal expressions}
For an application of the new modified Chen-Nester expressions
consider gravity theory.  With $\kappa=8\pi{}G$, let
\begin{equation}
q{}\rightarrow\Gamma^{\alpha}{}_{\beta}, \quad
p{}\rightarrow{}\frac{1}{2\kappa}\eta_{\alpha}{}^{\beta}.
\end{equation}
Here $\Gamma$ is the connection one form and
$\eta^{\alpha\beta}=*(\theta^{\alpha}\wedge\theta^{\beta})$ where
$\theta^{\alpha}$ is the coframe.  For gravity following
\cite{Nester 1, Nester 2}, rewrite (\ref{8Feb2006}) as
\begin{equation}\label{10eFeb2006}
2\kappa{\cal{}B}_{c_{1},c_{2}}(N)
=2\kappa{\cal{}B}_{p}(N)+c_{1}i_{N}\Delta{}\Gamma^{\alpha}{}_{\beta}
\wedge\Delta\eta_{\alpha}{}^{\beta}
-c_{2}\Delta\Gamma^{\alpha}{}_{\beta}
\wedge{}i_{N}\Delta\eta_{\alpha}{}^{\beta},
\end{equation}
where
\begin{equation}\label{10dFeb2006}
2\kappa{\cal{}B}_{p}(N)=\Delta\Gamma^{\alpha}{}_{\beta}
\wedge{}i_{N}\eta_{\alpha}{}^{\beta}
+\overline{D}_{\beta}\overline{N}^{\alpha}\Delta\eta_{\alpha}{}^{\beta}.
\end{equation}
Since we are interested in the energy-momentum components, the $DN$
terms in (\ref{10dFeb2006}) can be ignored (i.e. we can presume
$\overline{DN}=0$). Rewriting (\ref{10dFeb2006})
\begin{equation}\label{10fFeb2006}
2\kappa{\cal{}B}_{p}(N)
=\Delta\Gamma^{\alpha}{}_{\beta}\wedge{}i_{N}\eta_{\alpha}{}^{\beta},
\end{equation}
Using (\ref{10fFeb2006}), rewrite (\ref{10eFeb2006})
\begin{equation}\label{8Mar2006}
2\kappa{\cal{}B}_{c_{1},c_{2}}(N)
=\Delta\Gamma^{\alpha}{}_{\beta}\wedge{}i_{N}\eta_{\alpha}{}^{\beta}
+c_{1}i_{N}\Delta{}\Gamma^{\alpha}{}_{\beta}\wedge\Delta\eta_{\alpha}{}^{\beta}
-c_{2}\Delta\Gamma^{\alpha}{}_{\beta}\wedge{}i_{N}\Delta\eta_{\alpha}{}^{\beta}.
\end{equation}
For all $c_{1}$ and $c_{2}$, this expression will give good values
at the spatial infinity limit.  We can find preferred values for
$c_{1}$ and $c_{2}$ by considering the small region limit.\\

Consider first using Riemann normal coordinates and the adapted
orthonormal frames. As mentioned in~\cite{So}, if we use orthonormal
frames and the appropriate reference
$(\overline{\Gamma}^{\alpha}{}_{\beta}=0)$ with
$N^{\alpha}={\rm{}constant}$ within matter we get the expected
material energy-momentum tensor. In vacuum to lowest non-vanishing
order we get the Bel-Robinson tensor with a positive coefficient
only for the case $c_{1}=c_{2}=0$.  As explained in more detail
in~\cite{Szabados}, we should get Bel-Robinson tensor because it
has positive energy.\\

On the other hand we can consider the same formal expression, but
using holonomic variables and reference so that
$\overline{\Gamma}^{\alpha}{}_{\beta}=0$,
$N^{\alpha}={\rm{}constant}$ in the holonomic frame.  Again to
lowest order inside matter we get the desired material limit.  In
vacuum, however, the basic term ${\cal{}B}_{p}(N)$ (it is just the
quasilocal expression of KBLB~\cite{KBL} in the limit it reduces to
the Freud superpotential which gives the Einstein pseudotensor)
gives the value $\frac{1}{18}(4B_{\alpha\beta\lambda\sigma}
-S_{\alpha\beta\lambda\sigma})x^{\lambda}x^{\sigma}$ as has long
been known~\cite{MTW}.  This is not so good because the vacuum
energy is not positive.  Can some choices of $c_{1}$, $c_{2}$ save
the day? The answer is yes.  Rewrite (\ref{10fFeb2006})
\begin{equation}
2\kappa{\cal{}B}_{p}(N)=\Gamma^{\alpha}{}_{\beta}\wedge{}i_{N}\eta_{\alpha}{}^{\beta}
=-\frac{1}{2}\sqrt{-g}N^{\alpha}U_{\alpha}{}^{[\mu\nu]}\epsilon_{\mu\nu},
\end{equation}
where the Freud superpotential is
\begin{equation}
U_{\alpha}{}^{[\mu\nu]}
=-\sqrt{-g}g^{\beta\sigma}\Gamma^{\tau}{}_{\lambda\beta}
\delta^{\lambda\mu\nu}_{\tau\sigma\alpha}.
\end{equation}
Consider the $c_{1}$ term in (\ref{8Mar2006})
\begin{equation}
i_{N}\Gamma^{\alpha}{}_{\beta}\wedge\Delta\eta_{\alpha}{}^{\beta}
=-\frac{1}{2}\sqrt{-g}N^{\alpha}(h^{\mu\pi}\Gamma^{\nu}{}_{\alpha\pi}
-h^{\nu\pi}\Gamma^{\mu}{}_{\alpha\pi})\epsilon_{\mu\nu},
\end{equation}
and the $c_{2}$ term
\begin{eqnarray}
\Gamma^{\alpha}{}_{\beta}\wedge{}i_{N}\Delta\eta_{\alpha}{}^{\beta}
=-\frac{1}{2}\sqrt{-g}N^{\alpha} \left\{
\begin{array}{cccc}
(h^{\mu\pi}\Gamma^{\nu}{}_{\alpha\pi}
-h^{\nu\pi}\Gamma^{\mu}{}_{\alpha\pi})~~~\\
-(\delta^{\mu}_{\alpha}h^{\pi\rho}\Gamma^{\nu}{}_{\pi\rho}
-\delta^{\nu}_{\alpha}h^{\pi\rho}\Gamma^{\mu}{}_{\pi\rho})\\
+(\delta^{\mu}_{\alpha}\Gamma^{\lambda}{}_{\lambda}{}^{\nu}
-\delta^{\nu}_{\alpha}\Gamma^{\lambda}{}_{\lambda}{}^{\mu})~~~~~~~~\\
\end{array}
\right\}\epsilon_{\mu\nu}.
\end{eqnarray}
Therefore (\ref{8Mar2006}) can be rewritten as
\begin{eqnarray}\label{10cFeb2006}
2\kappa{\cal{}B}_{c_{1},c_{2}}(N)
&=&-\frac{1}{2}N^{\alpha}U_{\alpha}{}^{[\mu\nu]}\epsilon_{\mu\nu}
+\frac{1}{2}c_{1}\sqrt{-g}N^{\alpha}(h^{\mu\pi}\Gamma^{\nu}{}_{\alpha\pi}
-h^{\nu\pi}\Gamma^{\mu}{}_{\alpha\pi})\epsilon_{\mu\nu}\nonumber\\
&{}&-\frac{1}{2}c_{2}\sqrt{-g}N^{\alpha}\left\{
\begin{array}{cccc}
(h^{\mu\pi}\Gamma^{\nu}{}_{\alpha\pi}
-h^{\nu\pi}\Gamma^{\mu}{}_{\alpha\pi})~~~\\
-(\delta^{\mu}_{\alpha}h^{\pi\rho}\Gamma^{\nu}{}_{\pi\rho}
-\delta^{\nu}_{\alpha}h^{\pi\rho}\Gamma^{\mu}{}_{\pi\rho})\\
+(\delta^{\mu}_{\alpha}\Gamma^{\lambda}{}_{\lambda}{}^{\nu}
-\delta^{\nu}_{\alpha}\Gamma^{\lambda}{}_{\lambda}{}^{\mu})~~~~~~~~\\
\end{array}
\right\}\epsilon_{\mu\nu}.
\end{eqnarray}
From now on, the weighting factor $\sqrt{-g}$ will be dropped for
convenience.  Using (\ref{10cFeb2006}), the pseudotensor can be
obtained as
\begin{eqnarray}
t_{\alpha}{}^{\mu}=\partial_{\nu}\left\{
\begin{array}{cccc}
-U_{\alpha}{}^{[\mu\nu]} +c_{1}(h^{\mu\pi}\Gamma^{\nu}{}_{\alpha\pi}
-h^{\nu\pi}\Gamma^{\mu}{}_{\alpha\pi})
~~~~~~~~~~~~~~~~~~~~~~~~~~~~~~~~~~~~~~~~~~~~~~~~\\
-c_{2}(h^{\mu\pi}\Gamma^{\nu}{}_{\alpha\pi}
-h^{\nu\pi}\Gamma^{\mu}{}_{\alpha\pi}
-\delta^{\mu}_{\alpha}h^{\pi\rho}\Gamma^{\nu}{}_{\pi\rho}
+\delta^{\nu}_{\alpha}h^{\pi\rho}\Gamma^{\mu}{}_{\pi\rho}
+\delta^{\mu}_{\alpha}\Gamma^{\lambda}{}_{\lambda}{}^{\nu}
-\delta^{\nu}_{\alpha}\Gamma^{\lambda}{}_{\lambda}{}^{\mu})
\end{array}\right\}.
\end{eqnarray}
Inside matter at the origin a short calculation gives
\begin{equation}
2\kappa{}t_{\alpha}{}^{\beta}(0)=2G_{\alpha}{}^{\beta}(0)
=2\kappa{}T_{\alpha}{}^{\beta}(0).
\end{equation}
Just what we expect from the equivalence principle.  In detail the
energy density inside matter at the origin, the zeroth order term is
\begin{equation}
{\cal{}E}=-t_{0}{}^{0}(0)=-\frac{G_{0}{}^{0}(0)}{\kappa}=-T_{0}{}^{0}(0)=\rho,
\end{equation}
where $\kappa=8\pi{}G$ and $\rho$ is the mass-energy density.  The
momentum density is
\begin{equation}
{\cal{}P}_{k}=-t^{0}{}_{k}=-\frac{G^{0}{}_{k}}{\kappa}=-T^{0}{}_{k}.
\end{equation}
At the origin in vacuum, the zeroth and the first derivative are
\begin{equation}
t_{\alpha}{}^{\beta}(0)=0=\partial_{\mu}t_{\alpha}{}^{\beta}(0).
\end{equation}
The first non-vanishing contribution appears at 2nd order.  The
non-vanishing second derivatives in vacuum at the origin, after a
little lengthy computation, are
\begin{equation}\label{16aFeb2006}
\partial^{2}{}_{\mu\nu}t_{\alpha\beta}(c_{1},c_{2})
=\frac{1}{9}\left\{(4+c_{1}-5c_{2})B_{\alpha\beta\mu\nu}
-(1-2c_{1}+c_{2})S_{\alpha\beta\mu\nu}
+(c_{1}-3c_{2})K_{\alpha\beta\mu\nu} \right\},
\end{equation}
where the Bel-Robinson tensor $B_{\alpha\beta\mu\nu}$, tensors
$S_{\alpha\beta\mu\nu}$ and $K_{\alpha\beta\mu\nu}$ are defined as
follows
\begin{eqnarray}
B_{\alpha\beta\mu\nu}
&:=&R_{\alpha\lambda\mu\sigma}R_{\beta}{}^{\lambda}{}_{\nu}{}^{\sigma}
+R_{\alpha\lambda\nu\sigma}R_{\beta}{}^{\lambda}{}_{\mu}{}^{\sigma}
-\frac{1}{8}g_{\alpha\beta}g_{\mu\nu}
R_{\lambda\sigma\rho\tau}R^{\lambda\sigma\rho\tau},\\
S_{\alpha\beta\mu\nu}&:=&
R_{\alpha\mu\lambda\sigma}R_{\beta\nu}{}^{\lambda\sigma}
+R_{\alpha\nu\lambda\sigma}R_{\beta\mu}{}^{\lambda\sigma}
+\frac{1}{4}g_{\alpha\beta}g_{\mu\nu}
R_{\lambda\sigma\rho\tau}R^{\lambda\sigma\rho\tau},\\
K_{\alpha\beta\mu\nu}&:=&R_{\alpha\lambda\beta\sigma}R_{\mu}{}^{\lambda}{}_{\nu}{}^{\sigma}
+R_{\alpha\lambda\beta\sigma}R_{\nu}{}^{\lambda}{}_{\mu}{}^{\sigma}
-\frac{3}{8}g_{\alpha\beta}g_{\mu\nu}
R_{\lambda\sigma\rho\tau}R^{\lambda\sigma\rho\tau}.
\end{eqnarray}
The Bel-Robinson tensor has many nice properties \cite{Deser}
including energy positivity \cite{Szabados}. Consider
(\ref{16aFeb2006}), when $(c_{1},c_{2})=(0,0)$, $(0,1)$, $(1,0)$ and
$(1,1)$, they are classified as the original Chen-Nester four
boundary expressions. The results from \cite{So} are
\begin{eqnarray}
\partial^{2}{}_{\mu\nu}t_{\alpha\beta}(0,0)
&=&\frac{1}{9}(4B_{\alpha\beta\mu\nu}-S_{\alpha\beta\mu\nu}),\\
\partial^{2}{}_{\mu\nu}t_{\alpha\beta}(0,1)
&=&-\frac{1}{9}(B_{\alpha\beta\mu\nu}+2S_{\alpha\beta\mu\nu}+3K_{\alpha\beta\mu\nu}),\\
\partial^{2}{}_{\mu\nu}t_{\alpha\beta}(1,0)
&=&\frac{1}{9}(5B_{\alpha\beta\mu\nu}+S_{\alpha\beta\mu\nu}+K_{\alpha\beta\mu\nu}),\\
\partial^{2}{}_{\mu\nu}t_{\alpha\beta}(1,1)
&=&-\frac{2}{9}K_{\alpha\beta\mu\nu}.
\end{eqnarray}
None are of the desired pure Bel-Robinson form.  The general form of
the Taylor expansion for the Chen-Nester four expressions in compact
form is
\begin{eqnarray}
t_{\alpha}{}^{\beta}(k_{1},k_{2})
=2G_{\alpha}{}^{\beta}+\frac{1}{18} \left\{
\begin{array}{cccc}
(4+k_{1}-5k_{2})B_{\alpha}{}^{\beta}{}_{\xi\kappa}\\
-(1-2k_{1}+k_{2})S_{\alpha}{}^{\beta}{}_{\xi\kappa}~~\\
+(k_{1}-3k_{2})K_{\alpha}{}^{\beta}{}_{\xi\kappa}~~~~~~\\
\end{array}
\right\} x^{\xi}x^{\kappa}+O({\rm{Ricci}},x)+O(x^{3}),
\end{eqnarray}
Consider (\ref{16aFeb2006}) again, we want the coefficients of
$S_{\alpha\beta\mu\nu}$ and $K_{\alpha\beta\mu\nu}$ to vanish.
Taking $(c_{1},c_{2})=(\frac{3}{5},\frac{1}{5})$ gives
\begin{equation}
\partial^{2}{}_{\mu\nu}t_{\alpha\beta}\left(\frac{3}{5},\frac{1}{5}\right)
=\frac{2}{5}B_{\alpha\beta\mu\nu}.
\end{equation}
This result is good, because it only contains the Bel-Robinson
tensor.  Consequently the small region energy will be positive. The
general form of the Taylor expansion of the expression is
\begin{equation}
t_{\alpha}{}^{\beta}\left(\frac{3}{5},\frac{1}{5}\right)
=2G_{\alpha}{}^{\beta}+\frac{1}{5}B_{\alpha}{}^{\beta}{}_{\xi\kappa}
x^{\xi}x^{\kappa}+O({\rm{Ricci}},x)+O(x^{3}).
\end{equation}
The corresponding four momentum within a small coordinate sphere is
\begin{equation}
P_{\mu}\left(\frac{3}{5},\frac{1}{5}\right)
=\frac{1}{2\kappa}\int\frac{1}{5} B^{0}{}_{\mu\xi\kappa}
x^{\xi}x^{\kappa}d^{3}x\nonumber\\
=-\frac{r^{5}}{300G}B_{\mu{}0l}{}^{l}
=-\frac{r^{5}}{300G}B_{\mu{}000},
\end{equation}
where $B_{\mu{}0l}{}^{l}=B_{\mu{}000}$, $P_{\mu}=(-E,P_{i})$ and
energy $E>0$.  Alternatively the four momentum can be written in
terms of the electric and magnetic parts as follows
\begin{eqnarray}
P_{\mu}\left(\frac{3}{5}{},\frac{1}{5}\right)=-\frac{r^{5}}{300G}(
E_{ab}E^{ab}+H_{ab}H^{ab},2\epsilon_{c}{}^{ab}E_{ad}H_{b}{}^{d}),
\end{eqnarray}
where $B_{\mu{}000}
=(E_{ab}E^{ab}+H_{ab}H^{ab},2\epsilon_{c}{}^{ab}E_{ad}H_{b}{}^{d})$,
here $E_{ab}=R_{0a0b}$,
$H_{ab}=\frac{1}{2}C_{0amn}\epsilon_{b}{}^{mn}$.\\

In an earlier work we have constructed a $10$ parameter class of new
superpotentials that give rise to pseudotensors which have positive
Bel-Robinson small vacuum limit \cite{So1,So2}.  But they all seemed
very artificial.  Our $c_{1}$, $c_{2}$ expressions are special
cases. However they, in contrast, have clear meanings in terms of
the boundary conditions in the Hamiltonian formalism. Here we found
a simple specific value for $c_{1}$, $c_{2}$ which produces the
desired positive vacuum result.  There is one unique holonomic
Hamiltonian boundary expression with this property.  It is not one
of the 4 previously considered quasilocal expression but rather a
certain specific combination which corresponds to fixing a special
combination of Dirichlet and Neumann type boundary conditions.
Obtained from the boundary term in the variation of the Hamiltonian,
namely
\begin{equation}
\oint\left(
\frac{3}{5}i_{N}\delta\Gamma^{\alpha}{}_{\beta}\wedge{}\Delta\eta_{\alpha}{}^{\beta}
-\frac{1}{5}\delta\Gamma^{\alpha}{}_{\beta}\wedge{}i_{N}\Delta\eta_{\alpha}{}^{\beta}
-\frac{2}{5}i_{N}\Delta\Gamma^{\alpha}{}_{\beta}\wedge\delta\eta_{\alpha}{}^{\beta}
+\frac{4}{5}\Delta\Gamma^{\alpha}{}_{\beta}\wedge{}i_{N}\delta\eta_{\alpha}{}^{\beta}
\right).
\end{equation}

\section{Conclusion}
Chen and Nester proposed the four boundary expressions for the
quasilocal quantities by using the covariant Hamiltonian formalism.
Two of them are unique and correspond to imposing boundary
conditions on a covariant combination, the other two are
non-covariant. Generalizing the latter two cases, one can make some
modification by a simple adjustment of the four boundary
expressions.  The basis is using their four expressions. Applying
the modified expressions to gravity, we found using holonomic frames
a  nice Bel-Robinson result in vacuum for certain specific $c_{1}$,
$c_{2}$.

\section*{Acknowledgments}
This work was supported by grant from National Science Council of
the Republic of China under the grant number NSC 94-2112-M-008-038.

\end{document}